\pdfoutput=1

\documentclass[10pt,a4pap]{article}

\usepackage{graphicx}
\usepackage{amsthm}
\usepackage[margin=1.55cm, columnsep=0.75cm]{geometry}
\usepackage{amsmath}%
\usepackage{mathtools}%
\usepackage{wasysym}%
\usepackage[ruled,vlined]{algorithm2e}
\usepackage{bm}

\usepackage{enumitem}
\usepackage{etoolbox}

\def\Underline{\setbox0\hbox\bgroup\let\\\endUnderline}
\def\endUnderline{\vphantom{y}\egroup\smash{\underline{\box0}}\\}
\def\|{\verb|}

\usepackage[numbers]{natbib}
\usepackage[superscript]{cite}
\makeatletter
\renewcommand\@biblabel[1]{[#1]}
\makeatother

\usepackage{amssymb}

\LinesNumbered

\graphicspath{{graphics/}}

\theoremstyle{plain} 
\newtheorem{theorem}{Theorem}[section] 

\newtheorem{lemma}[theorem]{Lemma}

\theoremstyle{definition} 
\newtheorem{definition}{Definition}[section]
\newtheorem{example}{Example}[section]

\AtBeginEnvironment{corollary}{\setlist[enumerate,1]{label=\arabic*,font=\upshape}}
\AtBeginEnvironment{theorem}{\setlist[enumerate,1]{label=\arabic*,font=\upshape}}
\AtBeginEnvironment{lemma}{\setlist[enumerate,1]{label=\arabic*,font=\upshape}}
\AtBeginEnvironment{definition}{\setlist[enumerate,1]{label=\arabic*,font=\upshape}}
\AtBeginEnvironment{example}{\setlist[enumerate,1]{font=\upshape}}

\usepackage{authblk}
\begin{document}
   
\title{Suffix tree-based linear algorithms for multiple prefixes, single suffix counting and listing problems\thanks{This research did not receive any specific grant from funding agencies in the public, commercial, or not-for-profit sectors.}}




\author[1,*]{Laurentius Leonard}
\author[1]{Ken Tanaka}
\affil[1]{Department of Information Sciences, Faculty of Science, Kanagawa University}
\affil[*]{email: \tt{r201803452eb@jindai.jp}}

\maketitle

\begin{abstract}
Given two strings $T$ and $S$ and a set of strings $P$, for each string $p \in P$, consider the unique substrings of $T$ that have $p$ as their prefix and $S$ as their suffix.
Two problems then come to mind; the first problem being the counting of such substrings, and the second problem being the problem of listing all such substrings.
In this paper, we describe linear-time, linear-space suffix tree-based algorithms for both problems.
More specifically, we describe an $O(|T| + |P|)$ time algorithm for the counting problem, and an $O(|T| + |P| + \#(ans))$ time algorithm for the listing problem, where $\#(ans)$ refers to the number of strings being listed in total, and $|P|$ refers to the total length of the strings in $P$.
We also consider the reversed version of the problems, where one prefix condition string and multiple suffix condition strings are given instead, and similarly describe linear-time, linear-space algorithms to solve them.
\end{abstract}

\providecommand{\keywords}[1]{\textbf{\textit{Keywords: }} #1}

\begin{keywords}
String algorithms, linear-time algorithms,
suffix trees,
prefixes,
suffixes
\end{keywords}

\section{Introduction}

Algorithms and data structures for string processing have long been an important topic of research, with applications in fields not limited to computer science, such as in computational molecular biology\cite{DanGusfield}.
In this paper, we give linear-time, linear-space algorithms for the problems described as follows. Given two strings $T$ and $S$ and a set of strings $P$, for each string $p \in P$, consider the unique substrings of $T$ that have $p$ as their prefix and $S$ as their suffix.
Two problems then come to mind; the first problem being the counting of such substrings, and the second problem being the problem of listing all such substrings.
We give an $O(|T| + |P|)$ time algorithm for the counting problem, and an $O(|T| + |P| + \#(ans))$ time algorithm for the listing problem, where $\#(ans)$ refers to the number of strings being listed in total, and $|P|$ refers to the total length of the strings in $P$.
We also consider the reversed version of the problem; given two strings $T$ and $P$ and a set of strings $S$, count and list for each $s \in S$ the unique substrings of $T$ that have $P$ as their prefix and $s$ as their suffix, and show that the counting and listing problems can be solved in $O(|T| + |S|)$ and $O(|T| + |S| + \#(ans))$ time respectively.

Various related problems have been studied in the past. 
For instance, Manber and Baeza-Yates presented a suffix array-based algorithm\cite{Manber1991}\cite{Manber1993} for the following problem: given strings $T$, $P$, and $Q$ and an integer $k$, find all occurrences of $P\Sigma^iQ$ in $T$ for $0 \leq i \leq k$, where $\Sigma^i$ represents $i$ arbitrary symbols. Their algorithm finds the number of occurrences in $O(\log n)$ time, and lists all the occurrences in $O(n^{1/4}+ R)$ time.
Their problem is similar to ours, as both are problems to count and list substrings in a text that meet prefix and suffix conditions specified by the input strings. The main differences are the existence of an upper limit on length $k$ for the matched strings and
the lack of uniqueness condition on matched occurrences.
In addition, while their algorithm is very efficient in regard to running time, it requires a quadratic $O(n(k + m))$ space.

Another related problem is the matching problem where, given a set of patterns that may include multiple variable length don't care symbols, 
we must determine whether at least one of the patterns occurs in a given text.
An $O((|t| + |P|) \log |P|)$ time algorithm for this problem was proposed by Kucherov and Rusinowitch\cite{Kucherov1997}.
Note that by simply putting a variable length don't care symbol between a pair of prefix and suffix strings, we can cover most of the occurrences that we look for in the problem of this paper, hence the relation.

As these examples demonstrate, problems similar to the ones in this paper have been studied in the past, but they may not be in linear order for both time and space complexity.
Our main contribution in this research is therefore in identifying and solving a new problem in both linear time and space complexity.

\section{Definitions and notations}

\subsection{Strings and sets}
Let $\Sigma$ be an alphabet of finite size. Then, we say that an element of the set $\Sigma^*$ is a \emph{string}.
For a string of length $n$, $a_0 a_1 \cdots a_{n-1}$ $(a_0, \cdots, a_{n-1}\in\Sigma)$,
we call $a_i$ its index $i$ character.

In this paper, we consider indices to start from 0, meaning the first character of a string is its index $0$ character, the index $1$ character refers to the second character, and the last character of a string of length $n$
is the index $n-1$ character. Given a string $S$, we denote its length by $|S|$. Whenever we attach a terminal symbol ($\$$) to its end, the string's length includes that symbol as well. 
Given a set of strings $set$, we denote the sum of the length of its elements by $|set|$, while the number of its elements is written as $\#(set)$.

For a string $T$, we denote its index $i$ character by $T_i$, and its substring that starts from index $s$ and ends at index $e$ inclusive by $T_s \cdots T_e$.
Thus, string $T$ can also be expressed as $T_0 \cdots T_{|T|-1}$.

In addition, for two strings $a, b$, we define the following boolean functions.

\begin{equation}
pre(a, b) = \begin{cases} true &\mbox{if $a$ has $b$ as its prefix,} \\
false &\mbox{if $a$ does not have $b$ as its prefix.} \\
\end{cases}
\end{equation}

\begin{equation}
suf(a, b) = \begin{cases} true &\mbox{if $a$ has $b$ as its suffix,} \\
false &\mbox{if $a$ does not have $b$ as its suffix.} \\
\end{cases}
\end{equation}

\begin{equation}
sub(a, b) = \begin{cases} true &\mbox{if $a$ is a substring of $b$,} \\
false &\mbox{if $a$ is not a substring of $b$.} \\
\end{cases}
\end{equation}

\subsection{The problems}
We define the problems being considered in this paper below.

\begin{definition}[Multiple prefixes, single suffix counting and listing problem]
\label{DefProblem}
Given $2$ strings $T$ and $S$, and a set of $n$ strings $P = \{p_0, \cdots, p_{n-1}\}$,
let $ans_i$ denote the set of unique substrings of $T$ that have $p_i$ as their prefix and $S$ as their suffix.

\begin{itemize}
\item[] \textbf{Mutiple prefixes, single suffix counting problem.} For each 0 $\leq i \leq n-1$, find $\#(ans_i)$.
\item[] \textbf{Mutiple prefixes, single suffix listing problem.} For each 0 $\leq i \leq n-1$, find and list the elements of $ans_i$.
\end{itemize}
\end{definition}

Similarly, we define the reversed version of the problems, where a single prefix and multiple suffixes are given, as follows.

\begin{definition}[Single prefix, multiple suffixes counting and listing problem]
\label{DefProblem2}
Given $2$ strings $T$ and $P$, and a set of $n$ strings $S = \{s_0, \cdots, s_{n-1}\}$,
let $ans_i$ denote the set of unique substrings of $T$ that have $P$ as their prefix and $s_i$ as their suffix.

\begin{itemize}
\item[] \textbf{Single prefix, multiple suffixes counting problem.} For each 0 $\leq i \leq n-1$, find $\#(ans_i)$.
\item[] \textbf{Single prefix, multiple suffixes listing problem.} For each 0 $\leq i \leq n-1$ find and list the elements of $ans_i$.
\end{itemize}
\end{definition}

This paper excludes the empty string from the solution, which would otherwise only appear in the solution when both the prefix and suffix strings are empty strings. 

\subsection{Suffix trees}
Suffix tree is a data structure for string processing proposed by Weiner\cite{Weiner}.
Ukkonen's algorithm\cite{Ukkonen} constructs suffix trees in linear time, with suffix links added as auxiliary data.
In addition, 
for the suffix tree of a string $T$ constructed with Ukkonen's algorithm,
a \emph{pair of indices} $[s, e]$ is added to each edge of the suffix tree,
which represents the label of the edge, $T_s \cdots T_e$.

\begin{definition}[Suffix tree]
\label{DefSuffixTree}
We define the suffix tree of a string $T \in \Sigma^*$ as
the data structure obtained by
inserting all the suffixes of $T\$$ into
an initially empty compressed trie, where $T\$$ denotes the string $T$ concatenated with the terminal symbol $\$$, where  $\$\not\in\Sigma$.
\end{definition}

In this paper we consider the suffix trees constructed by a linear-memory construction algorithm, in particular Ukkonen's algorithm, which means the edges have pairs of indices that represent the label of the edge, as mentioned above.
Throughout this paper, we refer to such a pair of indices on an edge as the \emph{index-pair} of that edge.

A compressed trie is a trie
where, whenever a node has only one child node, the two nodes are merged, consequently allowing for edges with labels of length greater than $1$.
Compressed tries are also known by other names, such as radix tries, Patricia trees, as well as Patricia tries\cite{Szpankowski1990}.
Suffix trees are known to have many applications in the field of string algorithms, \cite{Grossi1993}\cite{DanGusfield}\cite{Pampapathi2006} and we note that other than suffix trees for single strings as defined above, suffix trees for multiple strings \cite{Inenaga2020Pointer}\cite{Takagi2019} and single tries \cite{Breslauer1998}\cite{Inenaga2021Towards} have also been studied.

While technically $T$ itself does not end with the terminal symbol, and only its depiction in the suffix tree does, throughout this paper we will treat the input string $T$ as $T\$$, i.e. as having a terminal symbol added to its end.
Note that 
for any string $x \in \Sigma^* \cup \{\$\}$, $sub(x, T) \wedge pre(x, p) \wedge suf(x, s) \Leftrightarrow sub(x, T\$) \wedge pre(x, p) \wedge suf(x, s)$ given $T, p, s \in \Sigma^*$,
meaning the substrings of $T$ that have $p$ as its prefix and $s$ as its suffix will always be the same as the substrings of $T\$$ that meet the same condition,
as substrings of $T\$$ that are not substrings of $T$ i.e. the substrings ending with the terminal symbol can never meet the condition since $s$ does not contain the terminal symbol.
Therefore, it is safe to treat $T$ and $T\$$ synonymously in the context of this paper's problems.
The only exception is when $s$ is the empty string, in which case all substrings of $T\$$ including the extra substrings not in $T$, i.e. the ones that end with $\$$ will have $s$ as suffix. To prevent this
corner case becoming a problem, we decided to make our auxiliary array, which we describe in a later subsection, treat these extra substrings as not having $s$ as suffix even when $s$ is the empty string.

\begin{figure}[h!]
  \centering
    \includegraphics[width=8cm]{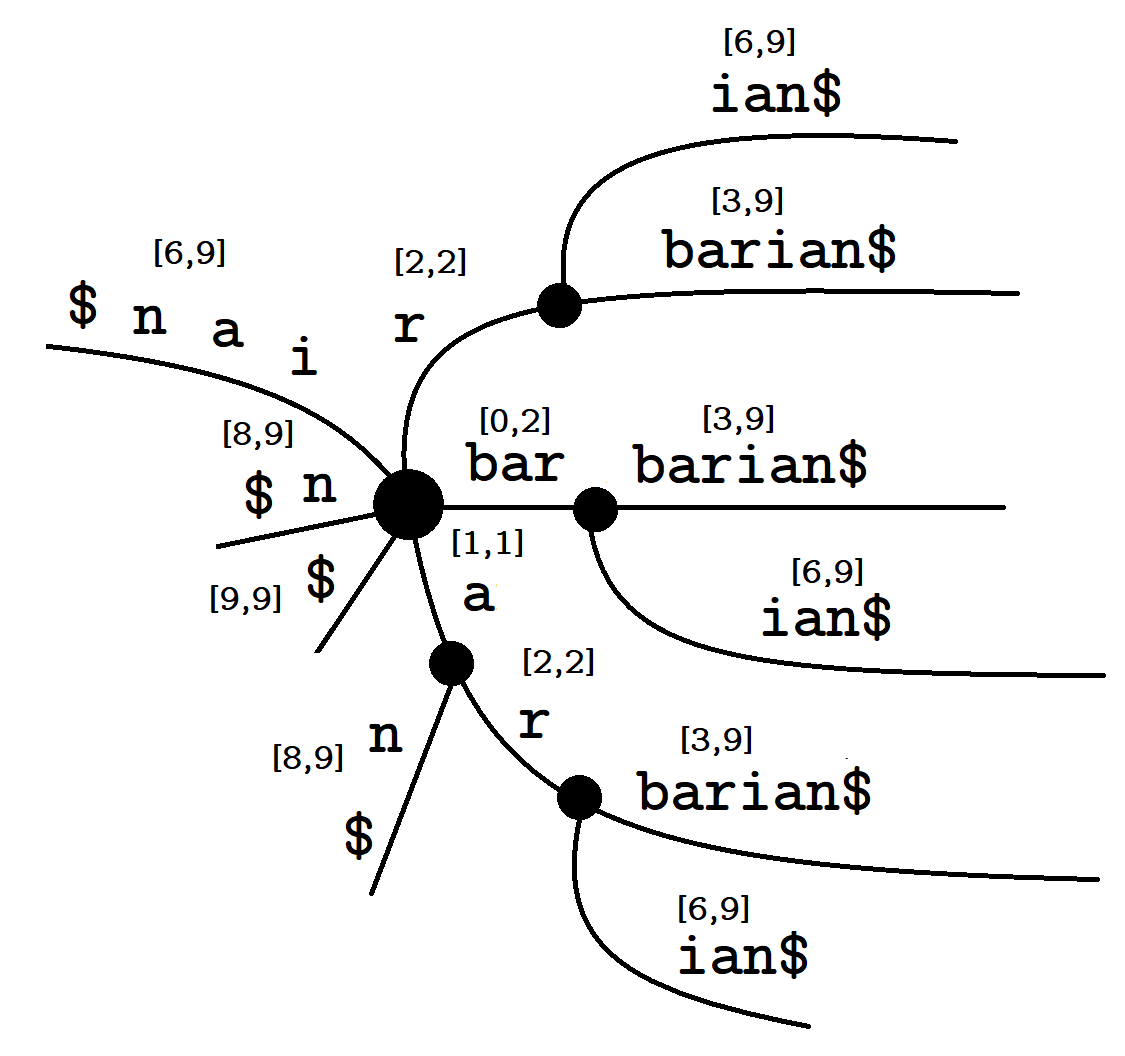}
    \caption{The suffix tree for $T=barbarian$.}
  \label{fig:bar}
\end{figure}

\begin{example}
\label{ExampleSuffixTree}
Figure \ref{fig:bar} shows the suffix tree for the string $T=barbarian\$$.
One can see that, for example, since $T_3 \cdots T_{9}=barian\$$, the three edges with the index-pair $[3, 9]$ has $barian\$$ as the label of their edges.

\end{example}

\subsection{Positions in suffix trees}

We denote by the \emph{position} of a string $X$ in the suffix tree of $T$ as the position reached by traversing $X$ from the root.
Similarly, the string of a given position is the label of the path from the root to that position.

\begin{figure}[h!]
  \centering
    \includegraphics[width=8cm]{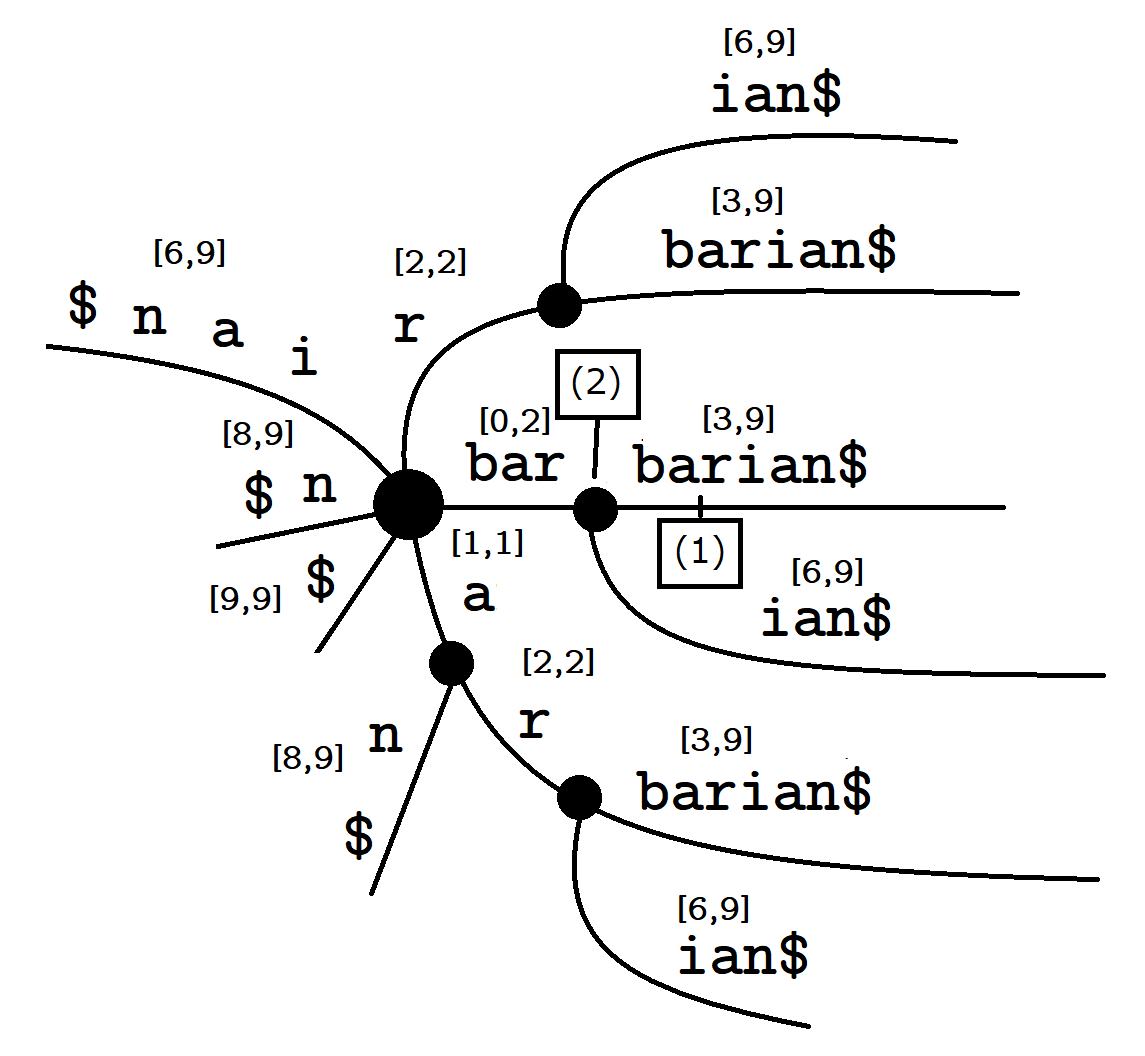} 
    \caption{Two positions in the suffix tree for $barbarian$.}
  \label{fig:bar2}
\end{figure}

\begin{example}
\label{ExampleSuffixTree2}
Figure \ref{fig:bar2} shows two positions in the suffix tree of $T=barbarian$.
Here, $(1)$ is the position of $barba$, and $(2)$ is the position of $bar$.
Notice that a position may be in the middle of an edge as with the case of $(1)$, or it may be on a node as shown with $(2)$.
Furthermore, note that although $bar$ occurs twice in $T$, it only has exactly one position in the suffix tree.
\end{example}

The lemmas below, that summarize well-known qualities of the suffix tree, immediately follow.

\begin{lemma}
\label{LemmaExistenceCondition}
The position of a string $X$ exists in the suffix tree of string $T$ if and only if $X$ is a substring of $T$.
\end{lemma}

\textbf{Proof. } 
By Definition \ref{DefSuffixTree}, the suffix tree of $T$ is a compressed trie with all of suffixes of $T$ inserted.
Therefore, any position in a suffix tree is either a position of a suffix of $T$, or a position in a path towards the position of a suffix of $T$, which in other words is the position of a prefix of a suffix of $T$, i.e. a substring of $T$.
Conversely, since all suffixes of $T$ are inserted into the tree, the positions of all suffixes of $T$ and all their prefixes must exist in the suffix tree. $square$

\begin{lemma}
\label{LemmaPosUniqueness}

The positions in the suffix tree of $T$ are in a one-to-one correspondence to the unique substrings of $T$.
\end{lemma}

\textbf{Proof.} By Lemma \ref{LemmaExistenceCondition}, there exists a position in the suffix tree for each of $T$'s substrings.
And since the suffix tree is a trie, exactly one position exists for each of $T$'s unique substrings. This implies a function that maps the set of unique substrings to the set of positions.
Furthermore, since the suffix tree is a tree structure, there is exactly one path from the root to any one position, meaning each position is only associated with exactly one of the unique substrings, that is, the path's label. Therefore, the function is a bijection. $square$

\begin{definition}[Depth of edges and nodes in a suffix tree]
\label{DefDepth}
The \emph{depth} of a \emph{node} in a suffix tree is the length of the path from the root to that node. In other words, the depth of a node is the length of the string associated with that node's position.
The \emph{depth} of an \emph{edge} in a suffix tree is the depth of the edge's source node.
\end{definition}


For an edge $edge$, consider the set of strings corresponding to the positions in the path from $edge$'s source node to the destination node, including the position of the destination node but not that of the source node.
We call this set of strings \emph{the strings associated with} $edge$, and we say that the strings in the set \emph{belong to} $edge$. Depending on the context, we will directly treat edges as the sets of strings associated with them.
For example, we denote the string $x$ belonging to $edge$ by $x \in edge$.
A formal definition is given below.

\begin{definition}[Strings associated with edges] 
\label{DefEdgeStrings}
Let $edge$ be an edge of depth $dep$, and $[s, e]$ be its index-pair.
Then, the set $\{T_{s-dep} \cdots T_i \mid s \leq i \leq e\}$ is the set of strings associated with $edge$.
\end{definition}

The above definition allows us to partition all the unique substrings of $T$ (except for the empty string) into different edges, such that each unique substring belongs to exactly one edge.
For a string $x$, we denote by $edge(x)$ the edge to which $x$ belongs.

\subsection{The SO array}

In solving the multiple prefixes, single suffix problems, we propose the suffix occurrence array, abbreviated as the $SO$ \emph{array}, an auxiliary array of integers that will help determine whether a given substring of $T$ has $S$ as its suffix.
This array is only dependent on the $T$ and $S$ input strings, and will be of great use in both the counting and listing problem.

$SO$ is an array of length $|T|$, where
for each occurrence of $S$ in $T$, we set the $SO$ value at the end-index of the occurrence in $T$ as $1$, while for each index in $T$ that is not an end-index of some occurrence of $S$, we set the $SO$ value of that index to $0$.
Additionally, as mentioned before, we will treat substrings that end with the terminal symbol $\$$ as not having $S$ as suffix, even in the case that it would be, i.e. when $S$ is the empty string.

A formal definition is given as follows.

\begin{definition}[The SO array]
\label{DefSO}
$SO$ is an array of integers of length $|T|$,
where the value for the element of index $i$ $(0 \leq i \leq |T|-1)$ is given by the following equation.

\begin{equation}
\label{equationSOCond}
SO[i] = \begin{cases} 1 &\mbox{if $i-|S|+1 \geq 0$ and $T_{i-|S|+1} \cdots T_i = S$ and $i \neq |T|-1$ ,} \\
0 &\mbox{otherwise.} \\
\end{cases}
\end{equation}
\end{definition}

\begin{lemma}
\label{LemmaSufCondition}
Any substring of $T$, $T_s \cdots T_e$ has $S$ as its suffix if and only if the length of $T_s \cdots T_e$ is at least $|S|$ and $SO[e]$=$1$.
Formally,
$suf(T_s \cdots T_e, S) \Leftrightarrow |T_s \cdots T_e| \geq |S| \wedge SO[e]=1$.
\end{lemma}

\textbf{Proof. }
The substrings that end with the terminal symbol $\$$ are treated as not having $S$ as their suffix for convenience as discussed before, and thus will have $SO$ value of $0$. This is realized
by the condition $i \neq |T|-1$ in Equation \ref{equationSOCond}. 

For any substring of $T$ that do not end with the terminal symbol, $sub = T_s \cdots T_e$,
when $|sub|<|S|$, $sub$ clearly cannot have $S$ as its suffix since $sub$ is shorter than $S$.

Consider the case where
$|sub|\geq|S|$.
Let $pre$ denote the prefix of $T$ that ends at $e$, i.e. $pre = T_0 \cdots T_e$.
Note that $pre$ has $sub$ as its suffix.
This implies that both $pre$ and $sub$ share the same last $|sub|$ letters.
Furthermore, since $|sub|\geq|S|$, both $pre$ and $sub$ share the last $|S|$ letters.
Therefore, when either of $pre$ or $sub$ has $S$ as its suffix, so does the other, and when one does not has $S$ as its suffix, neither does the other.

$pre$ has $S$ as its suffix if and only if there is an occurrence of $S$ in $T$ that ends at index $e$. By Definition \ref{DefSO}, this is also the necessary and sufficient condition for $SO[e]=1$.
Therefore, $sub$ has $S$ as its suffix if and only if $|sub|\geq|S|$ \emph{and} $SO[e]=1$. $square$

\section{The algorithms}

\subsection{The counting algorithms}

In this subsection, we describe our linear-time, linear-space algorithm that solves the \textbf{multiple prefixes, single suffix counting problem} defined in Definition \ref{DefProblem}.
We will further show that the reversed version of the problem, the \textbf{single prefix, multiple suffixes counting problem} can be similarly solved in linear order by applying reversal on the input strings.

Since for any $p_i \in P$, the unique substrings of $T$ that have $p_i$ as their prefix directly corresponds to the positions on the suffix tree reachable from $p_i$'s position as well as $p_i$'s position itself,
we can compute $\#(ans_i)$ by counting the positions reachable from $p_i$ whose strings have $S$ as their suffix.
Therefore, the key to solving the problem efficiently lies in the methodology of counting, for a segment of the suffix tree consisting of positions, how many of the positions
have $S$ as their suffix.

Our algorithm first partitions the suffix tree into segments before evaluating each of the segments.
To this end, we first define the keyword \emph{segment}, with which we express the parts of the suffix tree we partition it into.

\begin{definition}[Segments and their values] 
\label{DefEdgeSegment}
The $(a,s,e)$-$segment$ denotes the set of unique substrings of $T$ given by
$\{T_{a} \cdots T_i \mid s \leq i \leq e\}$.
The \emph{value} of a segment $seg$ is given by $\#(\{ x \mid x \in seg \wedge suf(x, S)\})$.

\end{definition}

For any edge $edge$ of depth $dep$ and index-pair $[s, e]$, the set of strings associated with that edge can be expressed as the $(s-dep,s,e)$-$segment$,
while any contiguous part of $edge$ consisting of the strings in $edge$ but with the shortest $x$ strings and the longest $y$ strings removed
can be expressed as the $(s-dep,s+x,e-y)$-$segment$.

Therefore, our definition of $segment$ allows us to express any edge of the suffix tree, or any contiguous part of it, as a $segment$.

Consequently, for any subtree of the suffix tree, whether its root position is within an edge or on a node, its positions can be partitioned into a set of $segments$, which each consist of positions belonging to different edges.


We further define the \emph{value of an edge} as the value of the $segment$ that expresses that edge, and the \emph{value of a node} as given below.


\begin{definition}[Value of a node]
The $value$ of a node $node$ is the sum of the values of edges that are reachable from $node$.
\end{definition}

In order to compute these values efficiently, we prepare two auxiliary arrays; the aforementioned $SO$ array and its cumulative sum array $CSO$, whose name is short for \emph{\textbf{c}umulative sum of \textbf{s}uffix \textbf{o}ccurrences}.

\begin{definition}[The CSO array]
\label{DefCSO}
$CSO$ is an array of integers of length $|T|$ whose values are the cumulative sum of the $SO$ array.
Formally, for each index $i$ $(0 \leq i \leq |T|-1)$, $CSO[i] = SO[0] + SO[1] + \cdots + SO[i]$.
\end{definition}

\begin{figure}[h!]
  \centering
    \includegraphics[width=4cm]{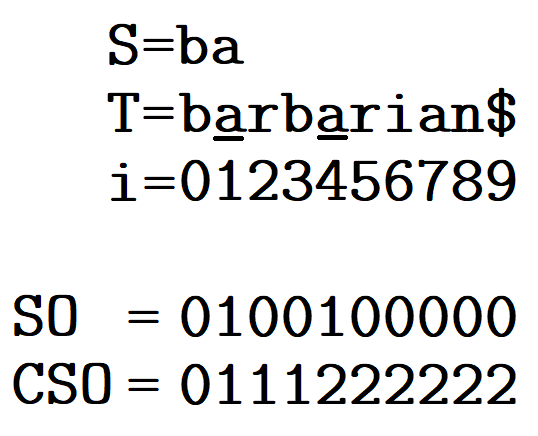}
    \caption{Computation for $SO$ and $CSO$ values with $T=barbarian$ and $S=ba$.}
  \label{fig:basocso}
\end{figure}

\begin{example}
\label{ExampleSOCSO}
Let $T=barbarian$, $S=ba$.
Then, $S=ba$ has exactly two occurrences on $T$; on indices $0$ and $3$.
As the two occurrences end at $1$ and $4$ respectively,
we have $SO[1] = SO[4] = 1$, while for all the other indices the values of $SO$ are 0.
Therefore, $SO = \{0, 1, 0, 0, 1, 0, 0, 0, 0, 0\}$ and $CSO = \{0, 1, 1, 1, 2, 2, 2, 2, 2, 2\}$.
Figure \ref{fig:basocso} shows the computation of these values.

\end{example}

Using the $CSO$ array allows us to compute the sum of $SO$ values within a range of indices in constant time via the function $CountSO$ given below.
\begin{definition}[CountSO]
\begin{equation}
\begin{multlined}
CountSO(s, e) = 
\begin{cases} 0 &\mbox{if $s<e$,} \\
 CSO[e]-CSO[s-1] &\mbox{if $s>0$,} \\
 CSO[e] &\mbox{if $s=0$.} \\
\end{cases}
\end{multlined}
\end{equation}
\end{definition}

Next, we show that the use of $CountSO$ allows for efficient computation of the values of $segments$ and nodes.

\begin{lemma}
Given the $CSO$ array, the value of any $segment$ denoted by the $(a, s, e)$-$segment$ 
is evaluated by \\
$CountSO(max(s, |S|+a-1), e)$ in constant time. 

\label{LemmaSegmentEvaluation}
\end{lemma}

\textbf{Proof.}
By definition, the value of the $(a, s, e)$-$segment$ is given by
$\#(  \{ x = T_{a} \cdots T_i \mid s \leq i \leq e \wedge suf(x, S)\}  )$.
This can be rewritten this using Lemma \ref{LemmaSufCondition} as follows.

\begin{equation}
\begin{aligned}
& \#(  \{ x = T_{a} \cdots T_i \mid s \leq i \leq e \wedge suf(x, S)\} \\
&= \#(  \{ x = T_{a} \cdots T_i \mid s \leq i \leq e \wedge |x| \geq |S| \wedge SO[i]=1\}) \\
&= \#(  \{ x = T_{a} \cdots T_i \mid s \leq i \leq e \wedge i-a+1 \geq |S| \wedge SO[i]=1\}) \\
&= \#(  \{ x = T_{a} \cdots T_i \mid s \leq i \leq e \wedge i \geq |S|+a-1 \wedge SO[i]=1\}) \\
&= \#(  \{ x = T_{a} \cdots T_i \mid max(s, |S|+a-1) \leq i \leq e \wedge SO[i]=1\}) \\
\end{aligned}
\end{equation}
Since $SO$ only take values of $0$ or $1$,
\begin{equation}
= \Sigma _{i=max(s, |S|+a-1)} ^e SO[i] \\
\end{equation}
Furthermore, since we assume the $CSO$ array is given, the sum of any given range of indices in $SO$ can be computed with $CountSO$.
Therefore, the value of the $(a, s, e)$-$segment$ can be computed by the function $EvaluateSegment(a, s, e)$ defined below.
\begin{equation}
EvaluateSegment(a, s, e) = CountSO(max(s, |S|+a-1), e)
\end{equation}

Thus, $CountSO(max(s, |S|+a-1), e)$ returns the value of any $segment$ written as the $(a,s,e)$-$segment$.
Since $CountSO$ takes constant time provided the $CSO$ array is given,
the value of any $segment$ can be computed in constant time when the $CSO$ array is computed in advance. $square$

\begin{lemma}
\label{LemmaTreeEval}
Given the $CSO$ array and the input string $T$,
all the values of nodes in the suffix tree of $T$ can be computed in $O(|T|)$ time.
\end{lemma}

\textbf{Proof. }
The number of nodes and edges in the suffix tree of $T$ is in the order of $O(|T|)$.\cite{DanGusfield}
Therefore, depth-first traversal of the suffix tree can be done in $O(|T|)$ time.
Since computing the value of any edge only takes constant time each,
and the value of each node can be computed by summing the values of its outgoing edges
and the values of destination nodes of these outgoing edges,
computing all the values only requires traversal of and constant time computation for each
node and edge, which adds up to $O(|T|)$.
A pseudocode for this computation is given by Algorithm \ref{AlgoEvalNode} below, which takes
the root of the suffix tree as its argument, and assigns the computed node values on the variable $eval$ of each node.


\begin{algorithm}[H]
\label{AlgoEvalNode}
\SetAlgoLined
\KwIn{Node $n$}
\KwResult{For each node in $n$'s subtree, their value is assigned to their $eval$ variable's value.} 
\KwOut{$n$'s value}
 n.eval := 0\;
 \ForEach{Edge $edge$ $\in$ outgoing edges of $n$}{
  n.eval += EvaluateEdge(edge)\;
  n.eval += EvaluateNode(edge's destination node)\;
 }
 \Return subTreeValue\;
 \caption{EvaluateNode(Node n)}
\end{algorithm}

Here $EvaluateEdge(edge)$ is a function that returns the value of the edge $edge$.
Since any $edge$ of depth $dep$ and index-pair $[s, e]$ can be expressed as the $(s-dep, s, e)$-$segment$,
by Lemma \ref{LemmaSegmentEvaluation}, we can define the function as below, which evaluates the edge value in constant time.
\begin{equation}
\label{eq:evedge}
EvaluateEdge(edge) = CountSO(max(s, |S| + s-dep-1), e)
\end{equation}


$EvaluateNode$ with the root as the argument starts a depth-first traversal done by recursively calling $EvaluateNode$
with each node as the argument exactly once throughout, during which the value for the node is computed by summing
the values of outgoing edges and their destination nodes. Since each edge only has one source node and each node only has up to one
parent node, and each node is only visited once, the value of each edge and node are only used in summing at most once throughout the whole algorithm.
Therefore, the algorithm ends in $O(|T|)$ time. $square$

Next, we show that after computing the node values in $O(|T|)$ time using the algorithm above,
the solution for each of $p \in P$ in the \textbf{multiple prefixes, single suffix counting problem} 
can be computed in $O(|p|)$ time.

\begin{lemma}
\label{LemmaEvalP}
Given the $CSO$ array and the values of all the nodes in $T$'s suffix tree,
for any given string $p$, the number of unique substrings of $T$ which have $p$ as their prefix and $S$ and their suffix can be computed in $O(|p|)$ time.
\end{lemma}

\textbf{Proof. } 
Denote by $reachable(x)$ the set of strings of positions reachable by the position of the argument string $x$.
Let $[s, e]$ be the index-pair of $edge(p)$, $dep$ the depth of $edge(p)$ and $destSubTree(edge(p))$ the
subtree rooted at the destination node of $edge(p)$.\footnote{As with edges, here we consider subtrees as sets of positions reachable by their root, but excluding the position of the root.}
Since the unique substrings of $T$ correspond to the positions reachable from $p$ as well as the position of $p$ itself, the solution is
the number of strings that have $S$ as their suffix in the set $\{p\}\, \cup \, reachable(p)$.

Partitioning the elements of $\{p\} \cup reachable(p)$ by separating those who belong to $edge(p)$ and those who do not, we obtain the following two sets:
\begin{equation}
\label{set12}
\begin{aligned}
& Set_1 = \{ x \in edge(p) \mid |x| \geq |p| \} \\
& Set_2 = destSubTree(edge(p))
\end{aligned}
\end{equation}

$Set_1$ can further be written as follows:
\begin{equation}
\label{set1seg}
\begin{aligned}
& Set_1 = \{ x \in edge(p) \mid |x| \geq |P| \} \\
&= \{ T_{s-dep}, \cdots, T_i \mid s \leq i \leq e \wedge i-s+dep+1 \geq |P| \} \\ 
&= \{ T_{s-dep}, \cdots, T_i \mid max(s, |P|+s-dep-1) \leq i \leq e \}  \\
&= (s-dep, max(s, |P|+s-dep-1), e)\text{-}segment
\end{aligned}
\end{equation}
Thus, we can express $Set_1$ as a segment.
The number of unique strings in $Set_1$ that have $S$ as their suffix can then
be obtained by evaluating the value of this segment, which by Lemma \ref{LemmaSegmentEvaluation} is obtained by the following:

\begin{equation}
CountSO(max(s, |P|+s-dep-1, |S|+s-dep-1), e)
\end{equation}

On the other hand, since $destSubTree(edge(p))$ is a set of positions consisting of the union of all the edges
reachable by the destination node of $edge(p)$, the number of unique strings in $Set_2$ that have $S$ as their suffix
is exactly the value of the destination node of $edge(p)$.

Thus, denoting the destination node of $edge(p)$ as $dest(edge(p))$,
the number of unique substrings of $T$ that have $p$ as their prefix and $S$ as their suffix is given by the following
when $edge(p)$ exists.

\begin{equation}
\label{ians}
CountSO(max(s, |p|+s-dep-1, |S|+s-dep-1), e) + dest(edge(p)).eval
\end{equation}

Here, traversal from the root to $p$ is required in order to find $edge(p)$ or determine that it does not exist. This takes $O(|p|)$ time.
When $edge(p)$ does not exist, we can safely conclude that there is no non-empty substring of $T$ that has $p$ as its prefix.
Otherwise, upon finding $edge(p)$, we immediately obtain $s$ and $e$ from its index-pair.
Since we assume that the $CSO$ is given, $CountSO$ takes constant time, and since the values of nodes are also given,
$dest(edge(p)).eval$ also takes constant time. Hence, given $CSO$ and the values of nodes, the value of Expression \ref{ians} can be found in $O(|p|)$ time. $square$

%

Combining the fact that we can obtain the solution for any of $p \in P$ in $O(|p|)$ time with the fact that the preprocessing takes $O(|T|)$ time, we obtain the following theorem.

\begin{theorem}
The solution for the \textbf{multiple prefixes, single suffix counting problem} can be found in $O(|T| + |P|)$ time.
\label{TheoremCount1}
\end{theorem}

\textbf{Proof.} The solution for the \textbf{multiple prefixes, single suffix counting problem} is computed by the following steps.

\begin{enumerate}
 \item Construct the suffix tree of $T$, using Ukkonen's algorithm.
 \item Construct the $SO$ and $CSO$ arrays.
 \item Do the function call $EvaluateNode(root)$. Here, $root$ refers to the root node of the suffix tree of $T$.
 \item For each $P$'s element $p_i \, (0 \leq i \leq n-1)$, traverse $p_i$. When the $edge(p_i)$ exists, compute $\#(ans_i)$ by evaluating Expression \ref{ians}, otherwise conclude that $\#(ans_i) = 0$.
\end{enumerate}

We show below that the total computation time for the above steps is in linear order.

\textbf{Step 1: } It is a known and proven fact that Ukkonen's algorithm constructs $T$'s suffix tree in $O(|T|)$ time.\cite{Ukkonen}.

\textbf{Step 2: }The construction of the $SO$ array starts with initializing it into an array of length $|T|$ where all the values are $0$.
After this, find all the occurrences of $S$ in $T$ using any pattern matching algorithm, and for each of the occurrences whose end-index is not $|T|-1$, set the SO value at the occurrence's end-index into $1$.
Whether we use classic algorithms such as the KMP or Boyer-Moore algorithm or by using the suffix tree\cite{DanGusfield}, finding all the occurrences can be done in $O(|T|)$ time.
As the construction of $CSO$ is simply taking the cumulative sum of $SO$ in ascending order of index, $CSO$ construction can clearly also be done in $O(|T|)$ time.

\textbf{Step 3: }By Lemma \ref{LemmaTreeEval}, Step 3 can be done in $O(|T|)$ time.

\textbf{Step 4: }By Lemma \ref{LemmaEvalP}, the value of Expression \ref{ians} is the solution $\#(ans_i)$, and can be evaluated in $O(|p|)$ time. Therefore, Step $4$ can be done in $O(|P|)$ time and finds all the solutions.

Therefore, the solution for the multiple prefixes, single suffix counting problem can be found in $O(|T| + |P|)$ time. $square$

\begin{example}
\label{exampleCount1}
Let $T=barbarian$, $P$'s elements $p_0=ba$, $p_1=bar$, $p_2=rb$, and $S=a$.
Then, the $SO$ and $CSO$ arrays and the value for each node is as shown on Figure \ref{fig:exval1}.

 \begin{figure}[h!]
  \centering
    \includegraphics[width=16cm]{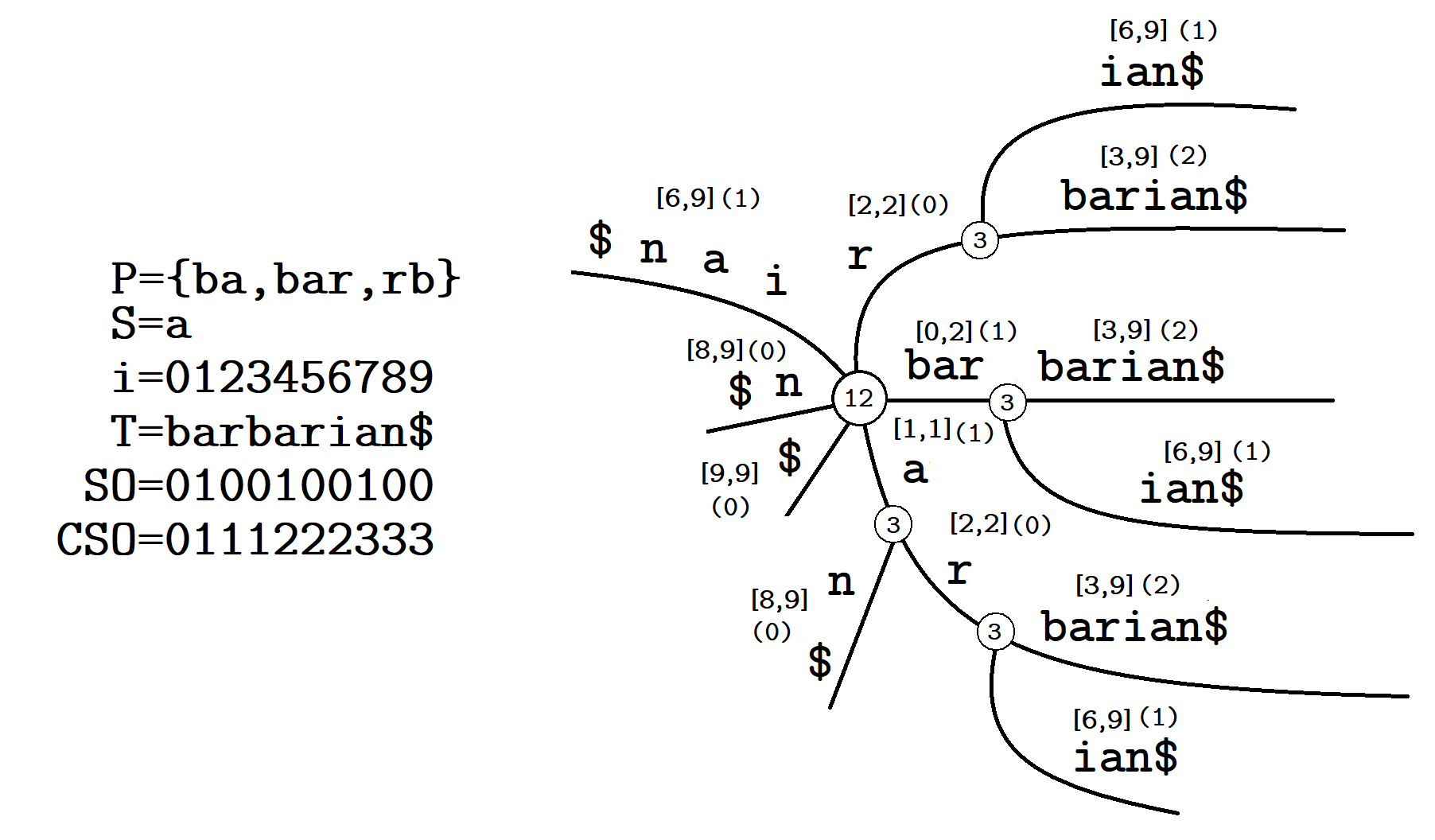}
    \caption{The suffix tree, SO and CSO arrays, and node values in the case of Example \ref{exampleCount1}. }
  \label{fig:exval1}
\end{figure}

For example, since the edge $edge(barba)$ has a depth of $3$ and index-pair $s=3, e=9$,
the value of $edge(barba)$ can be evaluated by Equation \ref{eq:evedge} as
$EvaluateEdge(edge(barba)) =$ $CountSO(3, 9) =$ $CSO[9] - CSO[2] =$ $3-1 = 2$.

Let us try evaluating $\#(ans_0)$ using Expression \ref{ians}.
From the problem input we know that $|p_0|=2, |S| = 1$,
and for the edge-dependent variables we simply look at the edge of $p_0$, $edge(p_0) = edge(ba)$,
and from its index-pair and depth values we get $s=0, e=2, dep=0$.
As for $dest(edge(p_0)).eval$, since all node values are already computed and stored at each node,
we look at the stored variable value and get $dest(edge(p_0)).eval = 3$.
Substituting these values into Expression \ref{ians}, we get
$\#(ans_0)=CountSO(max(0,1,0),2) + 3$ $= CountSO(1,2)+3 = 4$.
In actuality, $\#(ans_0)=\#(\{ ba, barba, barbaria, baria \}) = 4$.

\allowbreak
Similarly, in order to compute $\#(ans_1)$,
since $edge(bar)$ is the same edge as $edge(ba)$, we get the same values
for the variables that are dependent on the edge;
$s=0, e=2, dep=0, dest(edge(p_1)).eval = 3$.
The only difference from the previous example is the length of the $P$ element string, which is $|p_1|=3$.
Thus, substituting these values to Expression \ref{ians} we get
$\#(ans_1)=CountSO(max(0,2,0),2) + 3$ $= CountSO(2,2)+3 = 3$.
In actuality, $\#(ans_1)=\#(\{ barba$, $barbaria, baria \}) = 3$.

Next, we try computing $\#(ans_2)$.
The variable values in this case are $s=3, e=9, dep=1, |p_2|=2$.
Since the destination node of $edge(p_2)$ is a leaf node,
the destination node value in this case is $0$.
Therefore, by substitution into Expression \ref{ians} we get $\#(ans_2)=CountSO(max(3,3,2),9) + 0$ $= CountSO(3,9) = 2$.
In actuality, $\#(ans_2)=\#(\{ rba, rbaria \}) = 2$.
 \end{example}

Next, we show that the algorithm above can also be applied to solve the \textbf{single prefix, multiple suffixes counting problem} if we apply reversal on the input strings beforehand.

Let us denote the string obtained by reversing the string $x$ by $x^{-1}$.
For example, given $T = abc$, $T^{-1}=cba$.
Furthermore, for a set of strings $X$, we denote the set of strings consisting of the reversed strings of $X$'s elements by $X^{-1}$.
For example, given $X=\{ab,bc,cb\}$, we get $X^{-1}=\{ba,cb,bc\}$.


Given strings $T$ and $S$, and a set of strings $P = \{ p_0, \cdots, p_{n-1} \}$ as the input for the \textbf{multiple prefixes, single suffix counting problem},
let $Ans1(T, P, S)$ be a function that returns
the solution of the problem arranged in an array in the form of
$\{\#(ans_0), \#(ans_1), \cdots , \#(ans_{n-1})\}$. In other words, for any index $i$, the array's element of index $i$ is the solution for $P$'s element $p_i$.


Thus, the value of the index $i$ element of $Ans1(T, P, S)$ is as expressed below.
\begin{equation}
Ans1(T, P, S)[i] = \{ x \in \Sigma^* \mid sub(x, T) \wedge pre(x, p_i) \wedge suf(x, S) \}
\end{equation}

Conversely, given strings $T$ and $P$, and a set of strings $S = \{ s_0, \cdots, s_{n-1} \}$ as the input for the \textbf{single prefix, multiple suffixes counting problem},
let $Ans2(T,P,S)$ be a function that returns
the solution of the problem arranged in an array in the form of
$\{ans_0, \cdots ,ans_{n-1}\}$.


The value of the elements of $Ans2(T, P, S)$ thus follows the equation below.
\begin{equation}
Ans2(T,P,S)[i] = \{ x \in \Sigma^* \mid sub(x, T) \wedge pre(x, P) \wedge suf(x, s_i) \}
\end{equation}

\begin{theorem}
The solution for the single prefix, multiple suffixes counting problem can be found in $O(|T| + |S|)$ time.
\label{m1}
\end{theorem}

\textbf{Proof.} In general, for any strings $x$, $T$, $P$, $S$, the statement below holds.
\begin{equation}
\label{reversal}
sub(x, T) \wedge pre(x, P) \wedge suf(x, S) \Leftrightarrow sub(x^{-1}, T^{-1}), suf(x^{-1},P^{-1}), pre(x^{-1},S^{-1})
\end{equation}
Therefore, $Ans2(T,P,S)[i]$ can be rewritten as follows.

\begin{equation}
\begin{aligned}
& Ans2(T,P,S)[i] = \{ x \in \Sigma^* \mid sub(x, T) \wedge pre(x, P) \wedge suf(x, s_i) \} \\
& = \{x \in \Sigma^x \mid sub(x^{-1}, T^{-1}), pre(x^{-1}, s_i^{-1}), suf(x^{-1}, P^{-1}) \} \\
& = \{x \in \Sigma^x \mid x^{-1} \in Ans1(T^{-1},S^{-1},P^{-1})[i]) \}
\end{aligned}
\end{equation}
Thus, $Ans2(T,P,S)[i] = Ans1(T^{-1},S^{-1},P^{-1})[i]^{-1}$, 
which implies $\#(Ans2(T,P,S)[i]) = \#(Ans1(T^{-1},S^{-1},P^{-1})[i])$.

Therefore, the solution for the \textbf{single prefix, multiple suffixes counting problem} for input $T$, $P$, $S$
is equal to the solution for the \textbf{multiple prefixes, single suffix counting problem} for input $T^{-1},S^{-1},P^{-1}$.

$T^{-1},S^{-1},P^{-1}$ can be found in $O(|T| + |S|)$ time by reversing their original strings.\footnote{We only consider the case where $|P| \leq |T|$, since when $|P| > |T|$ all the solutions clearly become $0$.}
Combining this with the fact that, by Theorem \ref{TheoremCount1}, the solution for the \textbf{multiple prefixes, single suffix counting problem} for input $T^{-1},S^{-1},P^{-1}$ can be found in $O(|T^{-1}| + |S^{-1}|) = O(|T| + |S|)$ time,
we obtain the conclusion that the solution for \textbf{single prefix, multiple suffixes counting problem} can be found in $O(|T| + |S|))$ time. $square$



\subsection{The listing algorithms}
In this subsection, we describe the linear-time algorithms for the 
\textbf{multiple prefixes, single suffix listing problem} and the 
\textbf{single prefix, multiple suffixes listing problem}.


Similarly to how we constructed the $CSO$ array in the preprocessing phase to help us efficiently compute the number of strings in a segment that have $S$ as their suffix,
here we construct an array we call the $NextSO$ array as part of the preprocessing to help us find all the strings in a $segment$ that have $S$ as their suffix.


\begin{definition}[The NextSO array]
\label{DefNextSO}
The $NextSO$ array is an array of integers of length $|T|$, where the value for the element of index $i$ is given by Equation \ref{NextSOValue} below,
where $RemSO$ is a function that takes an integer and returns a set of strings, given by $RemSO(i) = \{j \in \mathbb{Z} \mid i < j \leq |T|-1 \wedge SO[j] = 1 \}$.
\begin{equation}
\label{NextSOValue}
NextSO[i] = \begin{cases}
min(RemSO(i)) &\mbox{$ ( \#(RemSO(i)) > 0  ) $} \\
-1 &\mbox{$( \#(RemSO(i)) = 0 )$} \\
\end{cases}
\end{equation}
\end{definition}

$NextSO$ can be obtained easily by the following algorithm.

\begin{algorithm}[H]
\caption{GetNextSO(int[] SO)}
\label{GetNextSO}
\SetAlgoLined
\KwIn{int[] SO}
\KwOut{NextSO}
int[] NextSO := new int[$|$T$|$]\; 
int j := -1\;

\For{$($int i := $|$T-1$|$; i $\geq$ 0; i$\,$-$\,$-$)$}{
NextSO[i] := j\;
\uIf{SO[i] = 1} 
{j := i\;}
}

\Return ans\;
\end{algorithm}

\bigskip
By making use of the $NextSO$ array, we can find the strings that have $S$ as their suffix in any given $segment$, as done by the algorithm below.

\begin{algorithm}[H]
\label{AlgGetSO}
 \caption{GetSO($a$, $s$, $e$)}
\SetAlgoLined
\KwIn{Three integers $a,s,e$ that express the $segment$ given as the $(a,s,e)$-$segment$}
\KwOut{The strings in the $(a,s,e)$-$segment$ that have $S$ as their suffix}
int $i := max(s, a+|S|-1)$\;
\uIf{$i \geq |T|$}{return\;}
 \uIf{$SO[i]$ = 0}{i := $NextSO[i]$\;} 
 
 \While{$i \leq e$ and $i \neq -1$}{
 Output the string $T_a \cdots T_i$\;
 $i$ := $NextSO[i]$\;
 }
\end{algorithm}
\bigskip


\begin{lemma}
\label{LemmaGetSO}
Assume that the arrays $SO$ and $NextSO$ are already constructed,
and let $occ$ denote the set of strings in $(a,s,e)$-$segment$ that have $S$ as their suffix.
Then, $GetSO(a,s,e)$ outputs all the elements of $occ$ in $O(\#(occ) + 1)$ time.
\end{lemma}

\textbf{Proof.}
The set $occ$ can be written as
$occ = \{T_a \cdots T_i \mid s \leq i \leq e \wedge suf(T_a \cdots T_i, S) \}$.
Using Lemma \ref{LemmaSufCondition}, we can further rewrite the equation above as follows.
\begin{equation}
\begin{aligned}
& occ = \{T_a \cdots T_i \mid s \leq i \leq e \wedge i-a+1 \geq |S| \wedge SO[i] = 1\}\\
&= \{T_a \cdots T_i \mid s \leq i \leq e \wedge i \geq a+|S|-1 \wedge SO[i] = 1\}\\
&= \{T_a \cdots T_i \mid max(s, a+|S|-1) \leq i \leq e \wedge SO[i] = 1\}
\end{aligned}
\end{equation}

Note that for all $occ$'s elements, the start-index $a$ is the same. The elements of $occ$ only differs in their end-index determined by the iterator $i$,
implying a one-to-one correspondence between the elements of $occ$ and their $i$ values.
Therefore, in order to find the elements of $occ$, it suffices to find all the indices $i$ that satisfy the condition $max(s, a+|S|-1) \leq i \leq e \wedge SO[i] = 1$.


In Algorithm \ref{AlgGetSO}, the variable $i$ iterates through all the indices that satisfy the condition in increasing order.
First, $i$ is initialized to the lower bound, i.e. the minimum index value that may satisfy the condition, $max(s, a+|S|-1)$ in line 1.
If this index happens to be $|T|$ or greater, then there may be no indices within range that satisfies the condition, so the algorithm terminates. This is done in lines 2-3.
Next, the algorithm checks whether this initial value of $i$ satisfies $SO[i]=1$ in line 4, if not, then line 5 of the algorithm assigns to $i$ the value of $NextSO[i]$, which by definition is the smallest index $i'$ greater than the current value of $i$ when such $i'$ exists, or $-1$ otherwise.
On the other hand, if the initial value of $i$ already satisfies $SO[i]=1$, the algorithm simply proceeds without changing the value of $i$.

Consequently, on the first time the algorithm reaches line 6, the value of $i$ is the minimum index that is at least $max(s, a+|S|-1)$ and satisfies $SO[i]=1$.
To list all the indices that satisfy our condition, it thus suffices to list them in increasing order by using $NextSO[i]$ in each iteration to identify the next minimum index $i'$ that satisfies $SO[i'] = 1$ and assigning this new value to $i$,
only halting (possibly before the first iteration) when $i$ is no longer within the condition's range $(i > e)$ or it has been determined that there are no more indices $i'$ that satisfy $SO[i']=1$, which is when $i$'s value has been assigned $-1$ from the previous $NextSO[i]$ value during line 5 or 8.

Thus, the algorithm iterates in increasing order all the indices $i$ that satisfy $max(s, a+|S|-1) \leq i \leq e \wedge SO[i] = 1$, n line 1 the value ofexactly the end-indices of the elements of $occ$. For all these indices, the corresponding element of $occ$ is outputted in line 7, consequently outputting all of $occ$'s elements in increasing order of their end-index throughout the algorithm.

Lines 1-5 clearly take constant time.
Line 6 is done exactly $\#(occ) + 1$ times, as lines 7-8 are done exactly $\#(occ)$ times, or once for every $occ$'s element to iterate through and output them, and line 6 must be done exactly one more time than lines 7-8.
Note that the outputting of  $T_a \cdots T_i$ in line 7 can be done in constant time, as it suffices to output the index-pair $[a, i]$ to express the substring, rather than the whole substring which would take linear time and space.
Therefore, Algorithm \ref{AlgGetSO} outputs all the elements of $\#(occ)$ in $O(\#(occ)+1)$ time. $square$

\begin{example}
\label{exampleGetSO}

Consider the suffix tree for $T=barbarian$ as shown in Figure \ref{fig:bar}.


$\mathbf{(1)}$
Let $S=ba$,
$SO=\{0,1,0,0,1,0,0,0,0,0\}$, and
$NextSO=\{1,4,4,4,-1,-1,-1,-1,-1,-1\}$.

Below we show the process of listing all the strings in $edge(ba)$ i.e. the edge in the direction of $b$ from the root, that have $S$ as their suffix, using $GetSO$.


As the edge $edge(ba)$ is expressed by the $(0,0,2)$-$segment$, make the function call $GetSO(0,0,2)$.
In line $1$, the value of $i$ is initialized to $max(0,1)=1$. Since $SO[i]=1$, line 5 is not executed.
In line $7$, $T_a\cdots T_1 = ba$ is outputted. Afterwards,
in line $8$ $i$'s value becomes $4$, after which the while loop terminates because of $i > e$.
The algorithm then ends, and we obtain just $ba$ as the output, reflecting the fact that among
the strings that belong to $edge(ba)$, $\{b, ba, bar\}$, only $ba$ has
$S=ba$ as its suffix.


$\mathbf{(2)}$
Let $S=a$,
$SO=\{0,1,0,0,1,0,0,1,0,0\}$, and
$NextSO=\{1,4,4,4,7,7,7,-1,-1,-1\}$.

Below we show the process of listing all the strings in $edge(rba)$ that have $S$ as their suffix.
As $edge(rba)$ is expressed by the $(2,3,9)$-$segment$, we make the function call $GetSO(2,3,9)$.
In line 1 the value of $i$ is initialized to $max(3,2)=3$. Since $SO[i]=0$, line 5 gets executed, setting $i$'s value to $4$.
Thus, during the first execution of line $7$, $T_2\cdots T_4 = rba$ is outputted. Then in line $8$, $i$'s value becomes $7$,
so during the next execution of line $7$ the string $T_2\cdots T_8 = rbaria$ is outputted. The next execution of line $8$ sets $i$'s value to $-1$,
ending the while loop.
The resulting output is thus $\{rba, rbaria\}$.

\end{example}

Next, we discuss how to deal with the multiple elements of $P$ in order to achieve an overall $O(|T| + |P|)$ time.
For each $p_i \in P$,
let us partition the substrings of $T$ that have $p_i$ as their prefix into two sets, $p_i.Set_1$ and $p_i.Set_2$, based on whether they belong to $edge(p_i)$ or not, similarly to what was done in Equation \ref{set12}.

\begin{equation}
\label{set12i}
\begin{aligned}
& p_i.Set_1 = \{ x \in edge(p_i) \mid |x| \geq |p_i| \} \\
& p_i.Set_2 = destSubTree(edge(p_i))
\end{aligned}
\end{equation}

As $p_i.Set_1$ can be expressed as a $segment$,
the elements that further have $S$ as their suffix can be listed
with the $GetSO$ function in $O( \#(p_i.Set_1 \cap ans_i) +1)$ time.


On the other hand, using the $GetSO$ function on
$p_i.Set_2$ requires us the first partition the set based on the edges they belong to
in order to obtain the $segments$ that make up $p_i.Set_2$.
Since the number of edges is in $O(|T|)$ order,
a total of $O(\#(p_i.Set_2 \cap ans_i)+|T|)$ time would be required to list all the $ans_i$ elements in $p_i.Set_2$,
which raises the total time of finding all the elements of $ans_i$ to $O(\#(ans_i) + |T|)$.
This would mean that the algorithm would take a quadratic total time of $O(n|T| + |P| + \#(ans))$ across all of $P$'s elements.


This rise of computation time is caused by the fact that in the evaluation of $p_i.Set_2$,
all the edges that make up the subtree is evaluated, regardless of whether these edges actually contain
any string that has $S$ as its suffix.
Below, we describe a scheme that 
takes away the need of taking $O(|T|)$ time to evaluate every $p_i.Set_2$.
by traversing each of the edge only once across the whole algorithm to bring down the total running time to $O(|T| + |P| + \#(ans))$.

\bigskip

For every node $u$ in the suffix tree,
assign a list of integers $u.prefixIds$.
This list contains the integer $i$ if and only if $p_i.Set_2$ consists of the subtree rooted at $u$.
Therefore, for each edge $edge$ in the suffix tree, $edge$ is included in $p_i.Set_2$ if and only if $i$ is included in the $prefixIds$ of $edge$'s source node or any of its ancestors.
This is to help keep track which edges are included in the $Set_2$ of which $P$ elements.

Outputting $p_i.Set_2 \cap ans_i$, or in other words the solutions of each $p_i$ that belong in their $Set_2$, into their respective output list $ans[i]$ can then be done by the following method:
Do a depth-first traversal starting at the root of the suffix tree.
During this, maintain a stack of integers $idStack$,
such that $idStack$ contains $i$ if and only if $p_i.Set_2$ contains the edge being traversed.

Then, when traversing edge $edge$ of index-pair $[s,e]$ and depth $dep$,
since this edge is included in $p_i.Set_2$ for all $i \in idStack$,
store the elements of the output of $GetSO(s-dep,s,e)$ to a temporary list if $idStack$ is not empty.
Afterwards, if the list is not empty, add each element of this list to $ans[i]$ for all $i \in idStack$.
If $idStack$ is empty, we do not call $GetSO$ so as to not waste computation time, as the output does not belong in any $ans_i$.
If the list is empty, we do not attempt to add its elements to $ans[i]$ for all $i \in idStack$ so as to not wastefully iterate through $idStack$.

Since all edges are eventually traversed, for all $p_i \in P$, the output of $GetSO$ for edges that make up $p_i.Set_2$, and those edges only, are added to $ans[i]$, accomplishing our goal of adding 
$p_i.Set_2 \cap ans_i$ into $ans[i]$ for each $p_i$.

$idStack$ can be maintained by first initializing it to an empty stack, then whenever the traversal reaches a node $u$, add all the elements of $u.prefixIds$ to $idStack$,
and after traversal of the node (along with its subtree) is finished, or in other words during backtrack of the node, pop $idStack$ exactly $\#(u.prefixIds)$ times, effectively removing the elements of $u.prefixIds$ from $idStack$.





\bigskip

Using all the ideas described above, we describe the linear-time algorithm for the \textbf{multiple prefixes, single suffix listing problem}.
The pseudocode for the algorithm is given below.


\begin{algorithm}[H]
\label{GetMultiPSingleS}
\caption{GetMultiPSingleS(string T, string[] P, string S)}
\SetAlgoLined
\KwIn{string T, string[] P, string S}
\KwOut{List[] ans}

Preprocessing(); \tcp{Construction of the suffix tree, SO and NextSO arrays}  
 
\For{$($int i := 0; i $<$ n; i++$)$}{
	Set1 := GetSet1(P[i])\;
	ans[i] := new List()\;
	ans[i].add(GetSO(Set1))\; \tcp{add the strings outputted in the GetSO call into ans[i]}
	destNode.prefixIds.add(i)\;
}

Stack$<$int$>$ idStack := \{\}\;
GetSet2(root)\;
\Return ans\;
\end{algorithm}

\begin{algorithm}[H]
\caption{GetSet2(Node node)}
\label{GetSet2}
\SetAlgoLined
\KwIn{Node node}
\KwResult{For all $p_i \in P$, all the elements of $p_i.Set_2 \cap ans_i$ is added to $ans[i]$}

  \ForEach{int $i$ $\in$ node.prefixIds}{
    idStack.add(i)\;}

    \ForEach{Edge $edge$ $\in$ node.edges}
    { 
    
     \If{idStack is not empty}
     {
     \tcp{Make a list consisting of the substrings outputted by GetSO against edge.}
      List edgeSO := GetSO(the $segment$ that expresses $edge$)\; 
      
     \If{edgeSO is not empty}
     {
      \ForEach{int $i \in$ idStack}
  	 {
  	   \tcp{Add all elements of edgeSO into ans[i].}
   	    ans[i].addAll(edgeSO)\; 
   	 }
     }
     }

      GetSet2(edge.dest)\;
    }

  \ForEach{int $i$ $\in$ node.prefixIds}{
    idStack.pop();
    }
\end{algorithm}
\bigskip

We start by describing the function $GetMultiPSingleS$ shown in Algorithm \ref{GetMultiPSingleS}.
$GetMultiPSingleS$ is, as the name implies, a function that returns the solution to the \textbf{multiple prefixes, single suffix listing problem}.
The algorithm takes the input for the problem; strings $T, S$ and an array of strings $P$,
and returns an array of list of strings $ans$ as the output,
where the list $ans[i]$ corresponds to $ans_i$ in Definition \ref{DefProblem},
meaning by the end of the algorithm $ans[i]$ will have the exact same elements as $ans_i$.


In line 1, a preprocessing function is called that constructs the suffix tree using Ukkonen's algorithm, and also constructs the arrays $SO$ and $NextSO$.

In lines 2-7, for each $p_i \in P$, the solution strings that belong in $Set_1$, that is, $p_i.Set_1 \cap ans_i$ are added to $ans[i]$,
and $i$ is added to the $prefixIds$ list of the root node of the subtree that makes up $p_i.Set_2$.



Firstly, in line 3, $GetSet1(P[i])$ is called to obtain the $segment$ that expresses $p_i.Set_1$.
Next, in line 4, the algorithm initializes $ans[i]$ to an empty list,
after which in line 5 the elements of $p_i.Set_1 \cap ans_i$ are obtained using $GetSO$, which are then added to $ans[i]$.
Then, in line 6, $i$ is added to the $prefixIds$ list of $edge(p_i)$'s destination node, which is the root node of the subtree.

Here, $GetSet1(P[i])$ is a function that returns the $segment$ that expresses $p_i.Set_1$,
and also finds the destination node of $edge(p_i)$, and assigns it to $destNode$, which is used in line 6.
Both the $segment$ that expresses $p_i.Set_1$ and the destination node of $edge(p_i)$ are found by traversing $p_i$.
Refer to Equation \ref{set1seg} for how to express $Set_1$ as a $segment$.

In line 8, the stack of integers $idStack$ is initialized,
before $GetSet2$ is called, after which the algorithm terminates.



\bigskip

Next, we describe the function $GetSet2$ as shown in Algorithm \ref{GetSet2}.
The purpose of this function is to add the elements of $p_i.Set_2 \cap ans_i$ to $ans[i]$ for each $p_i$, which takes care of the $Set_2$ partition,
whereas the $Set_1$ partition was taken care of in lines 2-7 of Algorithm \ref{GetMultiPSingleS}. This is done, as discussed before, by making use of
the stack $idStack$ and the $prefixIds$ list of each node in depth-first traversal.

The argument $node$ is the node being traversed in the depth-first traversal.
Upon visiting the node,  the elements of $node.prefixIds$ is first added in lines 1-3.
These exact elements are removed upon backtracking $node$, which is done in lines 15-17 after recursively traversing and backtracking $node$'s descendants in lines 4-14.

Note that because of the traversal order, the elements of $node.prefixIds$ must be at the top of the stack during backtracking, as any elements that might have been added
on top of these elements during subsequent recursive calls must have been popped during the backtrack phases of the same function calls where they were added, meaning
to remove the elements added in lines 1-3, we only need to pop the stack as many times as the number of elements that were added in lines 1-3 of the same function call.

The core part of the function, lines 4-14, 
traverses each of the outgoing edges of $node$, $edge$, by iterating through them. Before the recursive call in line 13 is performed,
if $p_i.Set_2$ includes $edge$ for some $p_i \in P$, or in other words if $idStack$ is not empty,
we need to add the strings in the $segment$ that have $S$ as their suffix to all such $p_i$.
The algorithm does this by first checking whether $idStack$ is not empty in line 5,
then in case that it is not, line 6 makes a list $edgeSO$ that contains all the strings in $edge$ that have $S$ as suffix, obtained by calling $GetSO$ against the $edge$ as a $segment$.
The elements of $edgeSO$ is then added to every $ans[i]$ for each $i \in idStack$ through lines 8-10, but line 7 of the algorithm will skip this if $edgeSO$ happens to be empty, in which case iterating through the elements $idStack$
would be a waste of computation time.
Doing this with all edges assures that the elements of $p_i.Set_2 \cap ans_i$ across all edges are all added to $ans[i]$ for all $p_i \in P$.





\begin{lemma}
\label{LemmaListMS}
$GetMultiPSingleS$ returns the solution for the \textbf{multiple prefixes, single suffix listing problem}.
\end{lemma}

\textbf{Proof.}
In lines 2-7 of $GetMultiPSingleS$,
for each $p_i$,
the elements of $p_i.Set_1 \cap ans_i$ are added to $ans[i]$.
In addition, $i$ is added to the $prefixIds$ of the root node of the subtree that makes up $p_i.Set_2$.



In the function $GetSet2$,
since the elements of $prefixIds$ of each node is added to $idStack$ upon visited and removed from $idStack$ upon backtracking,
during traversal of edge $edge$,
$idStack$ will contain exactly the integers $i$ where $p_i.Set_2$ includes $edge$.
Let $(a,s,e)$-$segment$ be the $segment$ that expresses $edge$.
Then, adding the output of $GetSO(a,s,e)$ to $ans[i]$ for each of $idStack$'s element $i$ as done in lines 5-12
means to add the elements of $ans_i \cap edge \cap p_i.Set2$ to $ans[i]$ for each $i$.
Since this is done for all the edges,
the elements of $ans_i \cap p_i.Set2$ are added to $ans[i]$ for each $0 \leq i \leq n-1$.


Thus, for all $p_i$, both the elements of $ans_i$ in $p_i.Set_1$ and  $p_i.Set_2$
are added into $ans[i]$. The output array of lists $ans$ is therefore the solution to the problem. $square$


\begin{example}
\label{exampleAlgo}
Below, we describe the execution of Algorithm \ref{GetMultiPSingleS} against the input
$T=barbarian$, $P$'s elements $p_0=ba$, $p_1=bar$, $p_2=a$, $p_3=ar$, and $S=ar$.


\begin{figure}[h!]
  \centering
    \includegraphics[width=16cm]{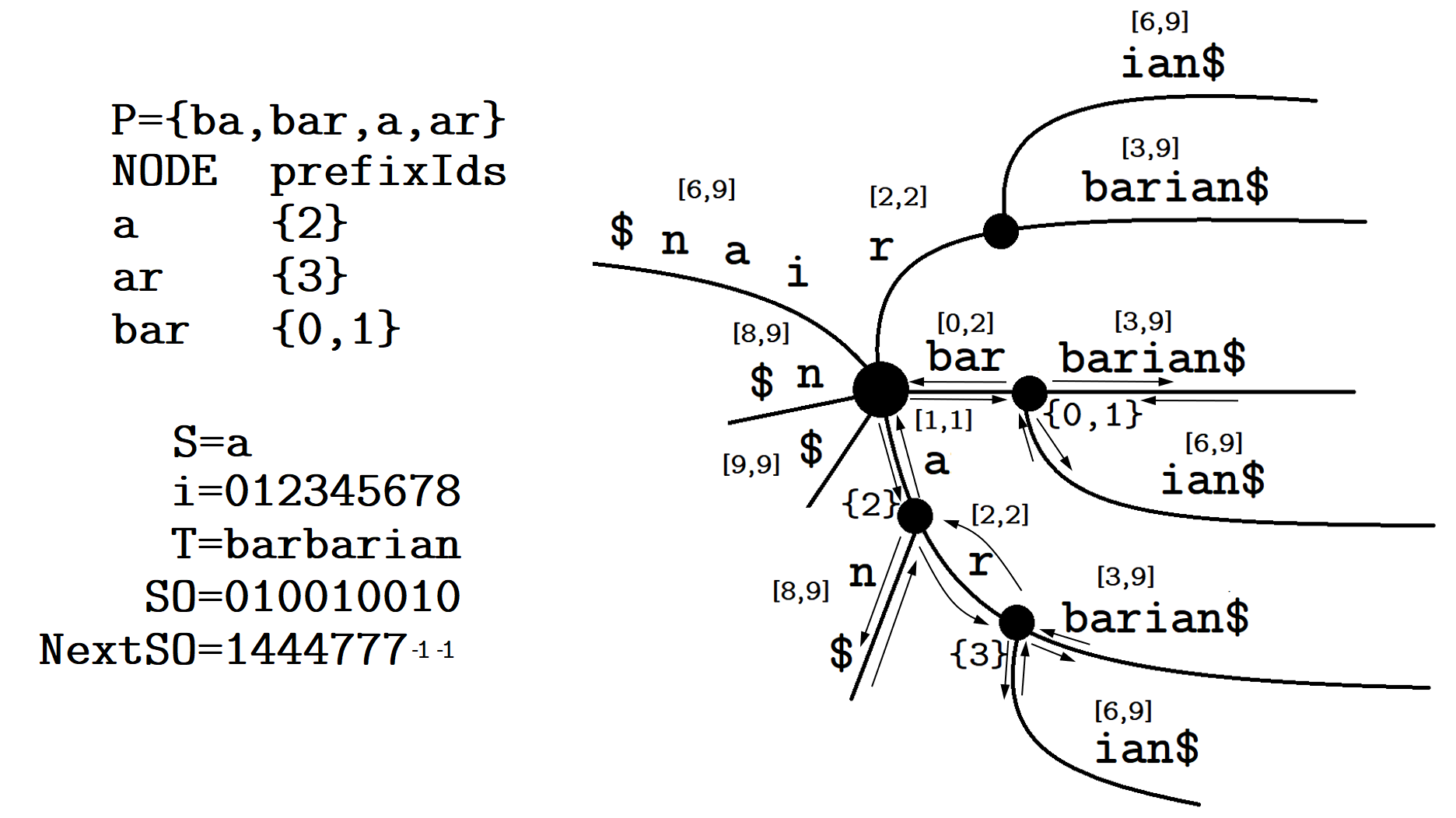}
    \caption{Execution of Algorithm \ref{GetMultiPSingleS} in Example \ref{exampleAlgo}}
  \label{fig:ex}
\end{figure}

The algorithm starts with the preprocessing in line 1. First, the suffix tree as shown in Figure \ref{fig:ex} is constructed using Ukkonen's algorithm.
Next, the $SO$ array in regard to $T$ and $S$ is constructed.
Since the three occurrences of $S=a$ in $T$ has end-indices $1, 4, 7$, 
$SO=\{0,1,0,0,1,0,0,1,0,0\}$ is obtained.
Afterwards, the $NextSO$ array is constructed as done in Algorithm \ref{GetNextSO},
which gives us $NextSO=\{1,4,4,4,7,7,7,-1,-1,-1\}$.

Next, we enter the for loop in lines 2-7.
For convenience, we will refer to the node reached by traversing the string $x$ as the $x$-node.



\begin{itemize}
\item During the $i=0$ iteration,
the algorithm first finds the $segment$ that expresses $p_0.Set_1$ in line 3.
Since $p_0=ba$, the algorithm does traversal from the root to the position of $ba$ and
finds $edge(ba)$,
and computes that the $segment$ that expresses $p_0.Set_1$ is the $(0,1,2)$-$segment$,
and also finds the destination of $edge(ba)$, the $bar$-node, and assigns it to $destNode$.
Moving on to line 4-5, $ans[0]$ is initialized to an empty list, to which
the output of $GetSO(0,1,2)$, which is $\{T_0\cdots T_1\}=\{ba\}$, is added.
Finally, in line 6, $0$ is added to the $prefixIds$ of the $bar$-node.

The same procedure is then executed for the iterations $i=1,2,3$, which we briefly describe below.



\item For the $i=1$ iteration,
we obtain $p_1.Set_1=(0,2,2)$-$segment$. The $GetSO$ execution for this segment outputs nothing, so $ans[1]$ remains empty.
As the destination node of $edge(bar)$ is the $bar$-node, $1$ is added to its $prefixIds$, which becomes $\{0, 1\}$.

\item For the $i=2$ iteration,
we obtain $p_2.Set_1=(1,1,1)$-$segment$. $GetSO$ outputs $\{a\}$ which is then added to $ans[2]$. 
As the destination node of $edge(a)$ is the $a$-node, $2$ is added to its $prefixIds$.

\item For the $i=3$ iteration,
we obtain $p_3.Set_1=(1,2,2)$-$segment$. $GetSO$ gives no output, so $ans[3]$ remains empty. 
As the destination node of $edge(ar)$ is the $ar$-node, $3$ is added to its $prefixIds$.



\end{itemize}

At this point, for all $p_i \in P$, the elements of $p_i.Set_1 \cap ans_i$ have been added to each of $ans[i]$,
and the necessary insertions of elements into $prefixIds$ lists have finished.
In order to also add the elements of $p_i.Set_2 \cap ans_i$ to each of $ans[i]$, the algorithm continues with the function call $GetSet2(root)$, which is done in line 9.


For the $GetSet2(root)$ call,
nothing is done in lines 1-3 as the $prefixIds$ of the root node is empty.
In line 4, the algorithm iterates through the outgoing edges of the root.
Let's assume the algorithm does this in dictionary order,
starting from the $a$-edge.
Lines 5-12 is skipped since $idStack$ is empty,
and the algorithm proceeds to calling $GetSet2(a$-$node)$ in line 13.

In the $GetSet2(a$-$node)$ call,
since the $a$-$node$'s $prefixIds$ contains the element $2$, lines 1-3 adds this to $idStack$,
resulting in $idStack = \{2\}$. This reflects the fact that the edges in $a$-$node$'s subtree
is included in $p_2.Set_2$ and no other $Set_2$'s.
Then we enter the for loop in lines 4-14, and iterate through all the outgoing edges of $a$-node,
starting with the $n$-edge which is expressed by the $(7,8,9)$-$segment$.
As $idStack$ is not empty,
$GetSO(7,8,9)$ is executed, but since its output is empty, $edgeSO$ is empty
and lines 8-10 are skipped; there is nothing to be added to $ans[2]$.
Since the destination node is a leaf node, there is no need to make a further call $GetSet2$ in line 13.
\footnote{It's only a matter of implementation whether or not to also make $GetSet2$ call against leaf nodes, such calls would terminate without causing any side effect as there would be no outgoing edges to iterate through in line 4. For this example we will skip the recursive call in line 13 when the destination node is a leaf node.}

The for loop in lines 4-14 continues with the iteration for the $r$-edge,
which also does nothing since the output of $GetSO(1,2,2)$ is also empty.
Line 13 makes the call $GetSet2(ar$-$node)$.

At lines 1-3 of the $GetSet2(ar$-$node)$ call,
$3$ is appended to $idStack$, resulting in $idStack = \{2,3\}$.
Lines 4-14 first iterate through the $b$-edge, expressed by the $(1,3,9)$-$segment$.
As $GetSO(1,3,9)$ outputs $\{arba, arbaria\}$, these are added to $ans[2]$ and $ans[3]$ in lines 6-11.
Next we enter the $i$-edge iteration, the edge expressed by the $(4,6,9)$-$segment$.
The output of $GetSO(4,6,9)$, $\{aria\}$, is added to $ans[2]$ and $ans[3]$ in lines 6-11.
This ends the traversal of the subtree rooted at the $ar$-node, and the algorithm proceeds to lines 15-17
where the idStack element $3$ that was added in this function call is removed, simply by popping $idStack$ as many times as the elements added in lines 1-3, in this case once.


The algorithm then continues the loop at line 4 of $GetSet2(a$-$node)$ call, with $idStack=\{2\}$. As there are no more outgoing edges to iterate, 
the traversal of the $a$-node subtree is finished, and the element $2$ is removed from $idStack$ courtesy of lines 15-17,
which brings us back to line 4 of $GetSet2(root)$ with $idStack$ empty.

Line 4 of $GetSet2(root)$ then proceeds to iterate the $b$-edge. Lines 5-12 is skipped as $idStack$ is empty,
and in line 13 $GetSet2(bar$-$node)$ is called.


In lines 1-3 of $GetSet2(bar$-$node)$, $0$ and $1$ are added to $idStack$.
Afterwards, lines 4-14 first iterate through the $b$-edge, expressed by the $(0,3,9)$-$segment$, and the output of $GetSO(0,3,9)$, $\{barba,barbaria\}$ is added to $ans[0]$ and $ans[1]$.
Next, we move to the iteration for the $i$-edge which is expressed by the $(0,6,9)$-$segment$, and the output of $GetSO(0,6,9)$, $\{baria\}$ is added to $ans[0]$ and $ans[1]$.
Lines 15-17 then removes $0$ and $1$ from $idStack$ by popping it twice,
after which we once again return to the $GetSet2(root)$ call with the $idStack$ empty.

After which, lines 4-14 of $GetSet2(root)$ iterates through the remaining outgoing edges, the $i,n,r$ edges, and traverses through all of them as well.
We skip these calls in this description, as the fact that there are no nodes with $prefixIds$ reachable through these edges mean that there are clearly no more elements to add into the solution arrays, as $idStack$ will remain empty throughout their traversal.
Thus, we obtain the final solution: $ans[0]=\{ba, barba, barbaria, baria\},$ $ans[1] = \{barba, barbaria, baria\},$ $ans[2] = \{a, arba, arbaria, aria\},$ $ans[3] = \{arba, arbaria, aria\}$.




\end{example}

\begin{theorem}

$GetMultiPSingleS$ computes and outputs the solution to the \textbf{multiple prefixes, single suffix listing problem} in $O(|T|+|P|+\#(ans_i))$ time.
\label{m2}

\end{theorem}

\textbf{Proof. }
By Lemma \ref{LemmaListMS}, $GetMultiPSingleS$ outputs the solution to the multiple prefixes, single suffix listing problem. 
We show below that this is accomplished in $O(|T|+|P|+\#(ans))$ time.

First, we consider the running time for the preprocessing done in line 1 of Algorithm \ref{GetMultiPSingleS}.
The suffix tree construction by Ukkonen's algorithm takes $O(|T|)$ time\cite{Ukkonen}.
Construction for the $SO$ array takes $O(|T|)$ time, as shown in the proof for Theorem \ref{TheoremCount1}.
Constructing $NextSO$ can be done in linear time by iterating through each index only once as shown in Algorithm \ref{GetNextSO}.
Thus, the preprocessing takes $O(|T|)$ time in total.


Next, we consider the running time for lines 2-7.
In line 3, we traverse from root to $p_i$'s position, which takes $O(|P|)$ time across all elements of $P$.
In line 5 $GetSO$ is called with the $segment$ that expresses $p_i.Set_1$ as the argument,
which takes $O(\#(p_i.Set_1 \cap ans_i) + 1)$ running time according to Lemma \ref{LemmaGetSO}.
Across all $p_i \in P$, lines 2-7 thus takes
$O(|P| + \Sigma_{i=0}^{n-1}\#(p_i.Set_1 \cap ans_i))$ time.


Next, we consider the running time for line 9, that is, the time complexity of $GetSet2(root)$.
Because lines 4-14 in Algorithm \ref{GetSet2} iterates through all the outgoing edges and recursively calls $GetSet2$ against the destination nodes,
$GetSet2$ is called against all the nodes in the suffix tree in depth-first order, for a total of $O(|T|)$ calls.

Since for each $p_i \in P$, only one node may contain $i$ in their $prefixIds$,
the total number of $prefixIds$ elements across all nodes is at most $\#(P)$.
Therefore, the total running time for lines 1-3 and 15-17 across all $GetSet2$ calls is $O(\#(P))$.

Lines 5-14 are done exactly once per edge in the suffix tree.
Recall that $idStack$ contain exactly the indices $i$ such that $edge \subseteq p_i.Set_2$,
and $GetSO$ in line 6 outputs $X = \{x \in edge \mid suf(x, S)\}$, which is also equal to $p_i.Set_2 \cap ans_i \cap edge$ for any $i \in idStack$.

In the case where $\#(idStack) > 0 \wedge \#(X) > 0$,
the conditions in lines 5 and 7 are both met,
and all of $x \in X$ are added to $p_i$ for all $i \in idStack$.
The adding of elements themselves takes constant time per elements added, which takes
$O(  \Sigma_{i = 0}^{n-1} \#(\{p_i.Set_2 \cap ans_i \cap edge\})  ) = O( \#(X)\#(idStack))$ time. Since this case assumes $\#(idStack) > 0 \wedge \#(X) > 0$,
the $O(\#(X) + 1)$ running time of $GetSO$ in line 6 and the $O(\#(idStack))$ time required to iterate through $idStack$ in line 8 does not raise this time complexity further for the total running time of lines 5-12.

In case where $\#(idStack) = 0 \vee \#(X) = 0$,
when $\#(idStack) = 0$, the condition in line 5 is not met and lines 5-12 thus takes $O(1)$ time, and when $\#(X) = 0$, $GetSO$ only takes $O(\#(X) + 1) = O(1)$ time, outputs nothing which causes $edgeSO$ to be empty
and thus skips lines 7-11, causing lines 5-12 to also only take $O(1)$ time.
In either case,  lines 5-13 terminates in $O(  \Sigma_{i = 0}^{n-1} \#(p_i.Set_2 \cap ans_i \cap edge) + 1 )$ time,
which across all edges adds up to $O(  \Sigma_{i = 0}^{n-1} \#(p_i.Set_2 \cap ans_i) + |T| )$ time as there are $O(|T|)$ edges.

$GetSet2(root)$ therefore takes $O(  \Sigma_{i = 0}^{n-1} \#(p_i.Set_2 \cap ans_i) + |T| + \#(P))$ time in total.
Adding all the time complexity calculated above together, we obtain the total time complexity for $GetMultiPSingleS$ as follows.



\begin{equation}
\begin{aligned}
& O(|T|+|P| + \Sigma_{i=0}^{n-1}\#(p_i.Set_1 \cap ans_i) + \Sigma_{i = 0}^{n-1} \#(p_i.Set_2 \cap ans_i))\\
=& O(|T|+|P|+\Sigma_{i=0}^{n-1}\#(ans_i))\\
=& O(|T|+|P|+\#(ans))
\end{aligned}
\end{equation}
Thus, we obtain the stated linear time complexity. $square$

Next, we show that the same algorithm can be applied to the \textbf{single prefix, multiple suffixes problem} with the use of string reversals.

\begin{theorem}
The solution to the \textbf{single prefix, multiple suffixes problem} can be found in $O(|T| + |S| + \#(ans))$ time.
\label{m3}
\end{theorem}

\textbf{Proof. }
Let us consider a function $GetSinglePMultiS(T, P, S)$ that returns the solution to the \textbf{single prefix, multiple suffixes listing problem}
as an array of list of strings, similarly to $GetMultiPSingleS$.


Then, by Equation \ref{reversal}, $Sol_1[i] = Sol_2[i]^{-1}$, the following equation holds for $0 \leq i \leq n-1$.

\begin{equation}
GetSinglePMultiS(T, P, S)[i] = GetMultiPSingleS(T^{-1},S^{-1},P^{-1})[i]^{-1}
\end{equation}

Therefore, the solution for the \textbf{single prefix, multiple suffixes listing problem} can be obtained
by reversing the strings in
$GetMultiPSingleS(T^{-1},S^{-1},P^{-1})$.
Computing $GetMultiPSingleS(T^{-1},S^{-1},P^{-1})^{-1}$
takes $O(|T|+|S|+\#(ans))$ by Theorem \ref{m2},
and reversing the strings in it takes $O(\#(ans))$ time assuming these substrings are expressed as pairs of indices.
Thus, the solution for the \textbf{single prefix, multiple suffixes problem} can be found in $O(|T| + |S| + \#(ans))$ time. $square$ 




\section{Conclusion}

In this paper, we proposed problems where multiple strings for prefix conditions and a single string for the suffix condition is given along with a text,
and showed linear-time algorithms to count and list the unique substrings of the text that match the conditions for each prefix-suffix condition pair.
In future work, we will consider related string processing problems and how data structures such as the suffix tree and the DAWG\cite{BlumerDawg} may be used to solve them efficiently.
One possible direction would be consider different versions of the current problem, such as when both the prefix and suffix strings are multiple, or when multiple texts are given.


\bibliographystyle{acm} 


\bibliography{ReferencesList}

\end{document}